\documentclass[prb,twocolumn]{revtex4}
\usepackage{times}
\usepackage{bbm}
\usepackage[dvips]{graphicx}
\usepackage{latexsym,amsmath,amssymb,bm,euscript}
\usepackage[dvips]{color}
\usepackage{multirow}
\usepackage{color}
\newcommand{\sgn}{\operatorname{sgn}}
\begin{document}

\title{Monte Carlo Study of a $U(1)\times U(1)$ Loop Model with Modular Invariance}
\date{\today}
\pacs{}

\author{Scott D. Geraedts}
\author{Olexei I. Motrunich}
\affiliation{Department of Physics, California Institute of Technology, Pasadena, California 91125, USA}

\begin{abstract}
We study a $U(1)\times U(1)$ system in (2+1)-dimensions with long-range interactions and mutual statistics. The model has the same form after the application of operations from the modular group, a property which we call modular invariance. 
Using the modular invariance of the model, we propose a possible phase diagram. We obtain a sign-free reformulation of the model and study it in Monte Carlo. This study confirms our proposed phase diagram. We use the modular invariance to analytically determine the current-current correlation functions and conductivities in all the phases in the diagram, as well as at special ``fixed'' points which are unchanged by an operation from the modular group. We numerically determine the order of the phase transitions, and find segments of second-order transitions. For the statistical interaction parameter $\theta=\pi$, these second-order transitions are evidence of a critical loop phase obtained when both loops are trying to condense simulataneously. We also measure the critical exponents of the second-order transitions.
\end{abstract}
\maketitle

\section{Introduction}

Models with statistical interactions can be used to describe a variety of interesting systems. In particular, quasiparticles in the Fractional Quantum Hall effect, as well as other fractionalized phases of spins and bosons, have such interactions.\cite{Read1989,Kitaev2003,SenthilFisher_Z2,Stern2008} Some models with statistical interactions contain a symmetry under the action of the modular group. This can simplify analytic study of these models. Several different such systems have been studied in the literature.
\cite{Fradkin_SL2Z,CardyRabinovici1982,Cardy1982,Shapere1989,Rey1991,Burgess2001,Lutken1993,Witten2003} 
In this work we define a model with this symmetry, which we call modular invariance, and study its properties both numerically and analytically. 

In this work, we study a $U(1)\times U(1)$ model in (2+1) dimensions with mutual statistical interactions. After introducing the model, we will explain what we mean when we say that it has modular invariance.
A general action for two species of $U(1)$ particles with mutual statistical interactions is given by the following action:
\begin{eqnarray}
S&=&\frac{1}{2}\sum_k \left[v_1(k)|\vec{J}_1(k)|^2+v_2(k)|\vec{J}_2(k)|^2\right]\nonumber\\
&+&\sum_k i\theta \vec{J}_1(-k)\cdot \vec{a}_2(k).
\label{action}
\end{eqnarray}
Here $\vec{J}_1$ and $\vec{J}_2$ represent conserved integer-valued currents residing on interpenetrating cubic lattices, and $\vec{\nabla} \cdot \vec{J}_1 = 0, \vec{\nabla} \cdot \vec{J}_2 = 0$. For brevity, the above action is defined in terms of Fourier components, where $v_1(k)$ and $v_2(k)$ are Fourier transforms of the intra-species interactions for species $J_1$ and $J_2$ respectively.  In the partition sum, a given current configuration obtains a phase factor $e^{i\theta}$ or $e^{-i\theta}$ for each cross-linking of the two loop systems, dependent on the relative orientation of the current loops. This is realized in the last term of Eq.~(\ref{action}), by including an auxiliary ``gauge field'' $\vec{a}_2$, whose flux encodes the $\vec{J}_2$ currents, $\vec{J}_2 = \vec{\nabla} \times \vec{a}_2$. As explained in our previous works,\cite{Loopy,short_range3} the model is precisely defined with periodic boundary conditions if, for all directions $\mu$,  $ J_{1\mu,{\rm tot}}\equiv\sum_r J_{1\mu}(r)=0$, and similarly for $J_2$.

\begin{figure}[t]
\includegraphics[angle=-90,width=\linewidth]{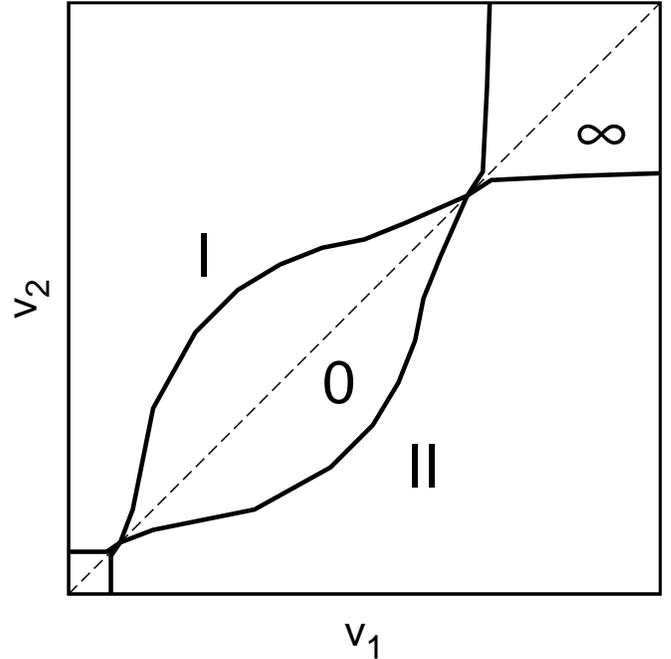}
\caption{Phase diagram for the model in Eq.~(\ref{action}), with fixed $\theta=2\pi/3$. The dashed line is the `symmetric' line where the potentials are equal, $v_1=v_2$, which is also assumed everywhere in the phase diagram in Fig.~\ref{SL2Z}. In phase ``$\infty$'' both $J$ variables are gapped, while they condense in phase ``0''. In phases I and II only one species is gapped. In the lower left corner different composite variables are gapped; here the structure can be significantly different for different values of $\theta$. }
\label{phase}
\end{figure}

Of some relevance to our study is the work of Fradkin and Kivelson.\cite{Fradkin_SL2Z}
Though several of the mathematical results in their work are applicable to our model, the model itself is different. In particular, Fradkin and Kivelson from the outset require a binding between different species, which is not present in our model and does not occur in our phase diagram.

Figure \ref{phase} shows a schematic of the phase diagram of the model in Eq.~(\ref{action}), for $\theta=\frac{2\pi}{3}$. In the remainder of this work, we use the following marginally long-ranged potential:
\begin{equation}
v_{1,2}(k)=\frac{2\pi g_{1,2}}{|\vec{f_k}|},
\label{V}
\end{equation}
where  $|\vec{f_k}|^2=\sum_\mu (2-2\cos{k_\mu})$, and $g_{1,2}$ are parameters describing the strength of the potentials. The $v(k)\sim\frac{1}{|k|}$ in momentum space is equivalent to a $1/r^2$ potential in real space. The main features of the phase diagram are controlled by the overall strength of the potentials, independent of the precise form (e.g.~in a previous work\cite{short_range3} we found a similar phase diagram for short-ranged potentials), while more detailed properties do depend on the range of potentials. The dashed line in the figure is the `symmetric' line, where $v_1=v_2$. 

When $v_1$ and $v_2$ are large, both $J_1$ and $J_2$ particles are gapped, and only small loop excitations are possible in these variables. We call this phase ``$\infty$''. If we decrease $v_1$ and $v_2$ along the symmetric line, the reduction in potential allows the $J_1$ and $J_2$ particles to condense, in the sense to be defined later; we call this phase ``0''. The labels of these phases will be explained in Sec.~\ref{sec:reformulations}. Having both particles condensed at the same time incurs some penalty due to the statistical interaction, and phase 0 exists only at intermediate couplings. As $v_1$ and $v_2$ are reduced still further, we can reach a phase where multiples of $J_1$ and $J_2$ (in particular, multiples of $n$ if $\theta=2\pi/n$) can condense. Roughly, such $n$-tuples of the current variables do not see a statistical interaction, the more precise meaning of this will be explained below. Off the symmetric line, we can access phases I and II where only one species of loop is condensed, and the other is gapped. The phase diagram is qualitatively similar for other values of $\theta$, with two exceptions. First, for $\theta=\pi$ phase 0 is not present. Instead there is a phase transition on the symmetric line, which separates phases I and II. An example of such a phase diagram can be seen in Ref.~\onlinecite{Loopy}. Second, for generic values of $\theta$ more phases exist at small $g$. These can be seen in vertical cuts in Figs.~\ref{SL2Z} and \ref{labels}, and will be explained in Sec.~\ref{sec:reformulations}.

The modular group is an infinite, non-abelian group, generated by two operations: duality (denoted by $S$) and periodicity (denoted by $T$). We call our action ``modular invariant'' because it has the same qualitative form after the application of these operations. The periodicity operation corresponds to shifting the statistical angle $\theta$ by an integer multiple of $2\pi$, and since 
the loop cross-linking number is an integer we can see that $e^{-S}$ for the action in Eq.~(\ref{action}) is unaffected by such shifts. In what follows we will find it useful to define $\eta\equiv\frac{\theta}{2\pi}$, and the complex number 
\begin{equation}
z=\eta+ig. 
\label{z}
\end{equation}
In this notation the action of such a shift by an integer $n$ is:
\begin{equation}
T^n:z\rightarrow z+n.
\label{shiftz}
\end{equation}

Duality corresponds to performing a well-known duality transform\cite{PolyakovBook, Peskin1978, Dasgupta1981, FisherLee1989, LeeFisher1989, artphoton,short_range3} on both species in the above action to obtain the following ``dual'' action:\cite{Ranged_Loops}
\begin{eqnarray}
S_{\rm dual}[\vec{Q}_1,\vec{Q}_2]&=&\frac{1}{2}\sum_k \left[ v_{1,{\rm dual}}|\vec{Q}_1(k)|^2+v_{2,{\rm dual}}|\vec{Q}_2(k)|^2\right]\nonumber\\
&+&\sum_k i\theta_{\rm dual}\vec{Q}_1(-k)\cdot \vec{a}_{Q_2}(k), \label{SQ}\\
v_{1/2,{\rm dual}}&=&\frac{(2\pi)^2v_{2/1}(k)}{|\vec{f_k}|^2v_1(k)v_2(k)+\theta^2},\nonumber\\
\theta_{\rm dual}&=&\frac{-(2\pi)^2\theta}{|\vec{f_k}|^2v_1(k)v_2(k)+\theta^2}.\nonumber
\end{eqnarray}
The $Q$ variables are dual to the $J$ variables and are also conserved integer-valued currents satisfying $\vec{\nabla}\cdot\vec{Q}_1=0$, $\vec{\nabla}\cdot\vec{Q}_2=0$. Under the exact duality $Q_{1,{\rm tot}}=Q_{2, {\rm tot}}=0$; $a_{Q_2}$ is an ``auxiliary'' field such that $\vec{Q}_2=\vec{\nabla}\times\vec{a}_{Q_2}$. If we think of the $J$ variables as boson number variables, the $Q$ variables are vortices in the boson phase variables. 

Let us use the long-ranged potential in Eq.~(\ref{V}), then we can see that on the symmetric line the action (\ref{SQ}) has the same form as (\ref{action}). The parameters in the original action transform under the duality in the following way:
\begin{equation}
g_{\rm dual}=\frac{g}{g^2+\eta^2}, \quad\eta_{\rm dual}=\frac{-\eta}{g^2+\eta^2}.
\label{gstar}
\end{equation}
In terms of the complex number $z$ we have 
\begin{equation}
S: z\rightarrow -1/z.
\label{dualz}
\end{equation}
Transformations $S$ and $T$ generate the modular group of transformations of the upper half of the complex plane $z$. Therefore with this choice of potential we say that the system is modular invariant. What happens here is that the statistical interaction can also be viewed as a marginally long-ranged interaction, and the duality operation preserves the form of such interactions. 
We chose the potential in Eq.~(\ref{V}) for analytical convenience because its form is preserved under duality, but it also corresponds to a three-dimensional Coulomb interaction between charged particles constrained to two spatial dimensions, and so we can apply this model to realistic systems. 

The phase diagram of a modular invariant system can be determined entirely from the properties of the modular group.\cite{Cardy1982,CardyRabinovici1982,Lutken1993,Fradkin_SL2Z} We will also use these modular transformations to characterize each phase of our model in terms of quasiparticles gapped in that phase. This will allow us to determine the behavior of current-current correlators and conductivities in each phase. 

Our numerical study also allows us to examine the critical properties of the system. All of the phase transitions in the modular invariant phase diagram can be mapped to each other under modular group operations. Therefore all such related phase transition points should have the same critical properties. We have found some phase transitions which are second-order, with continuously varying critical exponents. This is an example of a novel type of phase transition. We have also studied special points in the $\eta,g$ plane where three phases meet, and found these to be first-order. 

\section{Model and Monte Carlo Method}
\label{sec:MC}

The action in Eq.~(\ref{action}) has a sign problem, which must be eliminated if we are to study it in Monte Carlo. In order to do this, we dualize only one of the two loop species. In this work, we dualize the $J_1$ variables to get:\cite{short_range3}
\begin{eqnarray}
&&S[\vec{Q}_1,\vec{J}_2]=\frac{1}{2}\sum_k \Bigg[\frac{(2\pi)^2}{|\vec{f_k}|^2 v_1(k)}|\vec{Q}_1(k)|^2\label{Snum}\\
&&+\left(v_2(k)+\frac{\theta^2}{|\vec{f_k}|^2 v_1(k)}\right)|\vec{J}_2(k)|^2+\frac{4\pi\theta\vec{Q}_1(-k) \cdot \vec{J}_2(k)}{|\vec{f_k}|^2 v_1(k)}\Bigg].\nonumber
\end{eqnarray}
This is a sign-free classical statistical mechanics problem in terms of closed loops $Q_1$, $J_2$ and works for any $v_1$,$v_2$ and $\theta$ (Note that in Refs.~\onlinecite{Loopy,short_range3} we used a different sign-free reformulation that only works for a specific short-ranged $v_1$, $v_2$). 
In order to study the above action numerically, we write it in real space and use the potential from Eq.~(\ref{V}):
\begin{eqnarray}
&&S[\vec{Q}_1,\vec{J}_2]=\frac{1}{2}\sum_{r,r'}V(r-r') \times \bigg[ \frac{1}{g_1} \vec{Q}_1(r)\cdot\vec{Q}_1(r') \\
&&+\left(g_2+\frac{\eta^2}{g_1} \right)\vec{J}_2(r)\cdot\vec{J}_2(r')+ \frac{2\eta}{g_1}\vec{Q}_1(r)\cdot\vec{J}_2(r')\bigg]\nonumber,
\end{eqnarray}
\begin{equation}
V(r-r')=\frac{1}{\rm Vol}\sum_{k\neq0} \frac{2\pi }{|\vec{f_k}|}\cdot e^{i\vec{k}\cdot(\vec{r}-\vec{r}')},
\end{equation}
where ${\rm Vol}\equiv L^3$ is the volume of the system. In real space, $J_{2\mu}(r)$ is an integer-valued current on a link $r,r+\hat{\mu}$ of a cubic lattice. The variables $J_1$ are defined on a lattice dual to the lattice of the ${J}_2$, but after the duality procedure the ${Q}_{1}$ are integer-valued current variables on links of the same cubic lattice as the ${J}_2$.
We perform our simulations using the directed geometric worm algorithm. \cite{Sorensen} We attempt to produce worms in both the $Q_1$ and $J_2$ variables, while satisfying $Q_{1,{\rm tot}}=J_{2,{\rm tot}}=0$. 
In this work, we monitor ``internal energy per site'' $\epsilon\equiv S/$Vol, and compute ``specific heat'', defined as 
\begin{equation}
C=(\langle \epsilon^2\rangle-\langle \epsilon \rangle ^2) \times {\rm Vol}.
\end{equation}

In what follows, we will present all of our results in the $\vec{J}_1,\vec{J}_2$ language of Eq.~(\ref{action}). To study the behavior of these variables we wish to monitor current-current correlations, defined as:
\begin{equation}
C^{ab}_{\mu\nu}(k)\equiv\left \langle J_{a\mu}(k)J_{b\nu}(-k)\right \rangle,
\label{Cgen}
\end{equation}
where $a$ and $b$ are the loop species and $\mu$ and $\nu$ are directions; $J_{a\mu}(k) \equiv \frac{1}{\sqrt{\rm Vol}}\sum_{r} J_{a\mu}(r) e^{-i\vec{k} \cdot \vec{r}}$. 
We trivially have $C^{ba}_{\nu\mu}(k) = C^{ab}_{\mu\nu}(-k)$.
Because of the vanishing total current, we define the correlators at the smallest non-zero $k$; e.g., for $C^{aa}_{xx}$ we used $\vec{k}=(0,\frac{2\pi}{L},0)$ and $\vec{k}=(0,0,\frac{2\pi}{L})$. For simplicity, in this work we define $k$ to be in the $z$-direction, so that $k = (0, 0, k_z )$, and we only need to consider $\mu$, $\nu \in \{x, y\}$. From symmetry arguments \cite{short_range3} we know that $C^{aa}_{\mu\nu}$ is non-zero only if $\mu=\nu$, and $C^{12}_{\mu\nu}$ is non-zero only when $\mu\neq\nu$. Also, $C^{12}_{\mu\nu}=0$ when $\theta=\pi$.\cite{short_range3}

In our Monte Carlo we have access to the variables $\vec{J}_2$ and $\vec{Q}_1$. In order to monitor all correlators involving the $\vec{J}_1$ variables, we need to write $C^{11}_{\mu\mu}(k)$ and $C^{12}_{\mu\nu}(k)$ in terms of the correlators that we can measure: $C^{22}_{\mu\mu}(k)$, $\langle Q_{1\mu}(k)Q_{1\mu}(-k)\rangle$ and $\langle J_{2\mu}(k)Q_{1\mu}(-k)\rangle$. It is easy to argue that $\langle J_{2\mu}(k)Q_{1\mu}(-k)\rangle=\langle J_{2\mu}(-k)Q_{1\mu}(k)\rangle$, which are the only non-zero cross-correlators of $J_2$ and $Q_1$. 
To obtain expressions for $C^{11}_{\mu\mu}(k)$ and $C^{12}_{\mu\nu}(k)$ we can couple the original $\vec{J}$ variables to external probe fields $\vec{A}^{\rm ext}$ by adding the following terms to Eq.~(\ref{action}):
\begin{equation}
\delta S= i\sum_{k}\left[\vec{J}_1(-k)\cdot\vec{A}^{\rm ext}_1(k)+ \vec{J}_2(-k)\cdot\vec{A}^{\rm ext}_2(k)\right].
\label{Aextr}
\end{equation}
We carry the fields $\vec{A}^{\rm ext}_{1,2}$ through the duality procedure which leads to Eq.~(\ref{Snum}). By taking derivatives of the resulting partition sum with respect to the external fields, we can derive expressions for $C^{11}_{\mu\mu}$ and $C^{12}_{\mu\nu}$ in terms of correlators which we can measure: 
\begin{eqnarray}
&&C^{11}_{xx}(k)=\frac{1}{v_1(k)}-\frac{(2\pi)^2\langle Q_{1y}(k)Q_{1y}(-k)\rangle}{|\vec{f}_k|^2v_1(k)^2} \label{C11inQ}\\
&& -\frac{\theta^2C^{22}_{yy}(k)}{|\vec{f}_k|^2v_1(k)^2}-\frac{2(2\pi)\theta\langle Q_{1y}(k)J_{2y}(-k)\rangle}{|\vec{f}_k|^2v_1(k)^2},\nonumber\\
&&C^{12}_{xy}(k)=\frac{-\left[2\pi\langle Q_{1y}(k)J_{2y}(-k)\rangle+\theta C^{22}_{yy}(k)\right]}{2\sin{\frac{k_z}{2}}\cdot v_1(k)}\label{C12inQ}.
\end{eqnarray}

To be explicit, in the above equations we have set $\mu=x$, $\nu=y$. We note that on the symmetric line $v_1(k)=v_2(k)$ and so $C^{11}_{\mu\mu}(k)=C^{22}_{\mu\mu}(k)$. Whenever we present numerical data on the symmetric line, we have performed appropriate averages over both of these measurements and all directions to improve statistics. 

In order to determine the critical exponents of the model at various phase transitions, we will also monitor the derivatives of the correlation functions with respect to parameters in the potential. One option is to study derivatives with respect to $g$ (here we are working on the symmetric line where $g_1=g_2\equiv g$). However since $g$ controls marginally long-ranged interactions in our model, it is possible that universal properties, and in particular critical exponents, might depend on it.\cite{Ranged_Loops} To avoid possible difficulties in interpretation due to driving the transition while varying $g$, we have chosen to introduce a short-range interaction into the potential, so that 
\begin{equation}
v_1(k)=\frac{2\pi g}{|\vec{f_k}|}+t_1,\quad v_2(k)=\frac{2\pi g}{|\vec{f_k}|}+t_2,
\label{Vt}
\end{equation}
 where $t_1$ and $t_2$ are parameters controlling the strength of the additional short-range interaction. We can drive transitions by varying $t_1$ and $t_2$, with the expectation that critical indices will depend only on $g$ and $\eta$. We can fix $g$ at its critical value $g_{\rm crit}$, which we will find using our modular group analysis. We will extract critical exponents by taking derivatives with respect to $t_1$ and $t_2$, at $t_1=t_2=0$ and $g=g_{\rm crit}$. We will see in Sec.~\ref{sec:exponents} that we need symmetric and antisymmetric combinations of $t_1$ and $t_2$ to extract the critical exponents. We define $t_s$ to be the short-ranged parameter in the symmetric direction, and $t_a$ in the antisymmetric direction, which leads to $t_{1/2}=t_s \pm t_a$. When computing the derivative of a general expectation value $\langle O \rangle$, of an observable $O$, we use the following formula:
\begin{equation}
\left.\frac{\partial\langle O\rangle}{\partial t_{s/a}}\right|_{t_{s/a}=0}=
\left\langle O \right\rangle \left\langle \frac{\partial S}{\partial t_{s/a}}\right\rangle-
\left\langle O \frac{\partial S}{\partial t_{s/a}}\right\rangle.
\end{equation}
The action $S$ is the action given in Eq.~(\ref{Snum}), which is what is used in the Monte Carlo. 

The current-current correlations $C^{ab}_{\mu\nu}$ represent the response of the current $J_{a\mu}$ to an externally applied field $A^{\rm ext}_{b\nu}$. We can view our system with long-range interactions as having another, internal, gauge field, induced by the other currents in the system.\cite{MurthyShankarRMP, Herzog2007, Ranged_Loops}  In systems with short-range interactions, the quantity $C^{11}_{xx}(k)\cdot L$, with $k=k_{\rm min}\equiv(0,0,\frac{2\pi}{L})$ can be used to detect the phases of the system because it decreases with system size $L$ when the $J$ variables are gapped and increases when the $J$ variables are condensed. This allows the location of phase transitions to be determined by finding crossings of $C^{11}_{xx}(k_{\rm min})\cdot L$ at different $L$. However, the long-range interactions in our system prevent $C^{11}_{xx}(k_{\rm min})\cdot L$ from increasing when the $J$ variables condense,\cite{HoveSudbo2000,Ranged_Loops,short_range3,Herzog2007} so we cannot use crossings in this quantity to locate the phase transitions. To solve this problem, we study ``irreducible responses'', which measure the response of $J$ to the total field made up of $A^{\rm ext}$ and the internal field. These responses are related to the conductivities of the system. The derivation of these responses is given in Ref.~\onlinecite{short_range3}, and the result is the following equation for the conductivities:
\begin{eqnarray}
&&\mbox{\boldmath$\sigma$}=\frac{1}{|\vec{f}_k|}\mathbf{C}(\mathbbm{1}-\mathbf{VC})^{-1},
\quad \mbox{\boldmath$\sigma$}\equiv \left[ \begin{array}{cc} \sigma_{xx}^{11}(k) & \sigma_{xy}^{12}(k) \\ -\sigma_{xy}^{12}(k) & \sigma_{yy}^{22}(k) \end{array} \right]\nonumber\\
&&\mathbf C\equiv\left[ \begin{array}{cc} C^{11}_{xx}(k) & C^{12}_{xy}(k)\\ -C^{12}_{xy}(k) & C^{22}_{yy}(k) \end{array}\right],\\
&&\mathbf V\equiv \left[ \begin{array}{cc} v_1(k) & \frac{\theta}{2\sin(k_z/2)} \\ \frac{-\theta}{2\sin(k_z/2)} & v_2(k) \end{array} \right]\nonumber
.\nonumber
\end{eqnarray}
$\sigma_{xx}^{11}(k)$ relates the current induced in the $J_1$ variables in the $x$ direction to the total field in the $x$ direction, coupled to the same variables. $\sigma_{xy}^{12}(k)$ relates the current induced in the $J_1$ variables in the $x$ direction to the total field coupled to the $J_2$ variables in the $y$ direction. In Ref.~\onlinecite{short_range3} we showed that conductivities such as $\sigma_{xy}^{11}$ or $\sigma_{xx}^{12}$ are zero in our system. When we present numerical data we take appropriate averages over both species and all directions to improve statistics.
Unlike the current-current correlators, such $\sigma_{xx}^{11}(k_{\rm min})$ increase with $L$ in the phase where the $J_1$ and $J_2$ variables condense, even in the presence of long-range interactions, and therefore this quantity can be used to determine the phase transitions. 

\section{Phase diagram of the modular invariant model}
\label{sec:reformulations}
We now wish to use the modular invariance of our action to determine the phase diagram of the system with the $J_1\leftrightarrow J_2$ interchange symmetry, in the phase space defined by the intraspecies interaction $g$ and the statistical interaction $\eta$ ($\eta=\theta/2\pi$). 
To begin, consider the action in terms of the $J$ variables given in Eq.~(\ref{action}), using the potential in Eq.~(\ref{V}) with $g_1=g_2\equiv g$. The behavior of the $J$ variables is determined by the parameters $g$ and $\eta$. We know that as $g\rightarrow\infty$, the system will be in phase $\infty$, where the $J$ variables are gapped. As $g$ decreases, the $J$ variables will condense. To find the location of the phase transition, 
consider the action after the application of the duality operation $S$. This action is in terms of the $Q$ variables. 
Due to the fact that $V(r)\sim 1/r^2$, the $Q$ variables have the same kind of interaction as the original $J$ variables, with parameters $g_{\rm dual},\eta_{\rm dual}$, given by Eq.~(\ref{gstar}), instead of $g,\eta$. 

Consider the model at $\eta=0$. In this model, the two species of loops are decoupled, and $g_{\rm dual}=1/g$. There are two phases, one phase with the $J$ variables gapped (which we call phase $\infty$) and the other with the $Q$ variables gapped (phase 0). The phase transition between these two phases must occur at $g=g_{\rm dual}=1$. Such single loop models with long-ranged interactions were studied in Refs.~\onlinecite{Ranged_Loops} and \onlinecite{Kuklov2005}. 

Next, we can see from Eq.~(\ref{action}) that our model at phase space coordinates $(\eta,g)$ is mathematically equivalent to the model at $(-\eta,g)$, after making the change of variables $J_1\rightarrow -J_1$, while leaving $J_2$ unchanged. Therefore, away from $\eta=0$, we can use the equivalence between $\eta$ and $-\eta$ to see that the phase transition will again occur when $g=g_{\rm dual}$, which means that near $\eta=0$ the transition between phase $\infty$ and phase 0 occurs when $g^2+\eta^2=1$. We conjecture that this is the case for $-\frac{1}{2}<\eta<\frac{1}{2}$. We will see that this conjecture leads to a phase diagram which has the same properties after any operation by the modular group. This phase diagram is in agreement with our numerics. Therefore we believe that the conjecture is correct. We know that the phase diagram is periodic under integer shifts of $\eta$, so in the region $-\frac{1}{2}+n<\eta<\frac{1}{2}+n$ we expect a phase transition out of phase $\infty$ when $g^2+(\eta-n)^2=1$. Phase $\infty$ is located in the region of parameter space above these phase transitions. 

\begin{figure}[t]
\includegraphics[angle=-90,width=\linewidth]{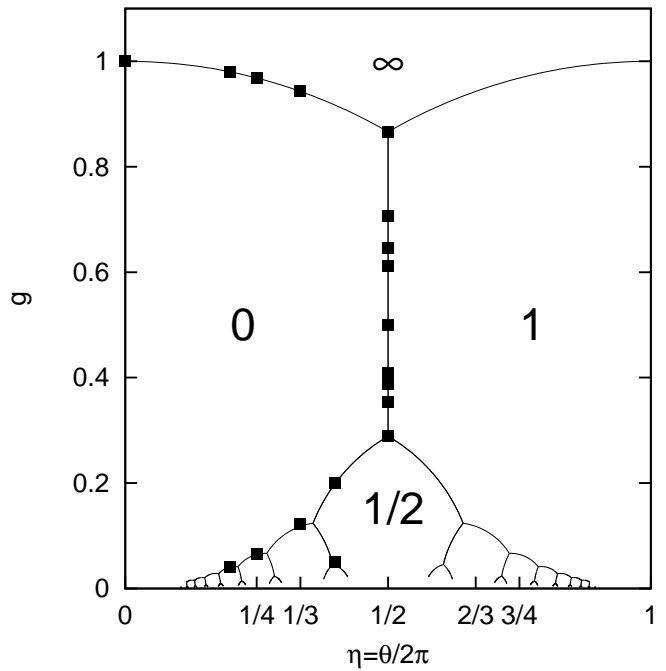}
\caption{The phase diagram for the ``symmetric'' model with $g_1=g_2=g$, for one period in $\eta$. The $J$ variables are gapped in phase $\infty$, and the (dual vortex) $Q$ variables are gapped in phase 0. In other phases, the gapped particles are linear combinations of $J$ and $Q$. Every phase can be mapped to phase $\infty$ by an operation in the modular group. Solid symbols show where the locations of phase transitions have been confirmed by our numerical study. There are infinitely many phase transitions in the model, so at small $g$ our diagram does not show every transition. The labels on the phases are explained in the text.} 
\label{SL2Z}
\end{figure}

We can now use the duality transform to determine that phase 0 is located in the region where $(\eta_{\rm dual}(\eta,g),g_{\rm dual}(\eta,g) )$ lie in the $\infty$ parameter region, i.e.~$g_{\rm dual}^2+(\eta_{\rm dual}-n)^2>1$, for some $n$ and $|\eta-n|\leq\frac{1}{2}$ (see also Fig.~\ref{SL2Z}). 

When deriving the extent of phase 0, we performed the following steps. First, we applied an operation of the modular group (specifically, duality) to the action in Eq.~(\ref{action}). This gave us an action in terms of new variables (specifically, we obtained the $Q$ variables). The new parameters in the action, $g_{\rm dual}$ and $\eta_{\rm dual}$, were functions of the original parameters $g$ and $\eta$. Note that both actions refer to the system at a single point on the phase diagram. By determining which $(\eta,g)$ gave $(\eta_{\rm dual},g_{\rm dual})$ in the $\infty$ parameter region, we were able to determine the extent of phase 0, where the $Q$ variables were gapped. 

We now want to generalize this procedure to everywhere in the phase diagram. This requires us to apply modular group operations more complicated than duality to the action in Eq.~(\ref{action}), and so we must determine the new phase space coordinates $(\tilde{\eta}$, $\tilde{g})$ that result from a given modular group operation. To do this we combine Eqs.~(\ref{shiftz}) and (\ref{dualz}) to get:\cite{Lutken1993,Fradkin_SL2Z}
\begin{equation}
\tilde{z}=\frac{az+b}{cz+d},
\label{zstar}
\end{equation}
where $a,b,c,d$ are integers and $ad-bc=1$. 
To find the $a,b,c,d$ that correspond to a given set of $S$ and $T$, we write them in matrix form:
\begin{equation}
\left [\begin{array}{cc} a & b \\ c& d\end{array} \right]
\end{equation}
and the operations can also be represented by matrices:
\begin{eqnarray}
&&S=\left[\begin{array}{cc} 0 & -1\\1& 0\end{array} \right],\\
&&T^n=\left[\begin{array}{cc} 1& n\\0& 1\end{array}\right].
\end{eqnarray}
We can find the $a,b,c,d$ that correspond to a given operation by multiplying these matrices. Such matrices where $\hat{A}$ and $-\hat{A}$ are identified make up the group $PSL(2,\mathbb{Z})$, which is equivalent to the modular group. 

We know that there is a phase transition at $g^2+\eta^2=1$, for $|\eta|\leq\frac{1}{2}$. If we apply a modular group operation, we will obtain an action in terms of variables with interactions $\tilde{g}$, $\tilde{\eta}$, and these variables will have a phase transition whenever $\tilde{g}^2+\tilde{\eta}^2=1,|\tilde{\eta}|\leq\frac{1}{2}$. Therefore we can find all of the phase transitions in the diagram by finding all the different values of $g, \eta$ which have this property for some modular operation. The resulting phase diagram is shown in Fig.~\ref{SL2Z}. We have only shown one period of the phase diagram, with $0\leq\eta\leq 1$, but the same structure repeats for all $\eta$. As the modular group is an infinite group, there are an infinite number of phase transitions, and so our diagram does not show all of the details at small $g$. 

The solid symbols in Fig.~\ref{SL2Z} show the locations of the phase transitions determined numerically. In a physical system $\eta$ must be fixed, and so we took data in sweeps at fixed $\eta$, varying $g$. This corresponds to vertical lines in the phase diagram. We determined the locations of phase transitions by observing peaks in the specific heat. We have also observed that $\sigma_{xx}^{11}(k_{\rm min})$ diverges with system size in phase 0 but decreases with system size in all of the phases neighboring phase 0 (this will be explained in Sec.~\ref{sec:correlations}). Therefore we were also able to use crossings in this quantity to locate the phase transitions.\cite{Ranged_Loops} 

Finally, using Eqs.~(\ref{C11inQ}) and (\ref{C12inQ}) we can determine that at the $\infty$-0 phase transition, where the $J$ variables and $Q$ variables see the same potential (but opposite statistical angle),
\begin{equation}
2\cdot[gC^{11}_{xx}(k)-\eta C^{12}_{xy}(k)\sgn{(k_z)}]=\frac{|\vec{f}_k|}{2\pi}.
\end{equation}
In addition, we find that in the thermodynamic limit the above quantity divided by $|f_k|$ approaches zero in phase $\infty$ and a different finite value in phase 0, and so we used crossings of this quantity as another way to find the location of the transition.

We have mentioned above that by applying an operation of the modular group we can express the original problem in terms of new loop variables. For a given phase, if we choose the modular group operation which gives $(\tilde{\eta},\tilde{g})$ in the $\infty$ parameter region, these loop variables will be gapped quasiparticles in that phase. In the following we will show how to determine the precise physical nature of the phase and these quasiparticles. Starting with Eq.~(\ref{action}), we perform a duality transformation on the $J_1$ variables to obtain Eq.~(\ref{Snum}). We then make the following substitutions:
\begin{eqnarray}
\vec{G}_1&=&c\vec{Q}_1-d\vec{J}_2.\label{subs1}\\
\vec{F}_2&=&a\vec{Q}_1-b\vec{J}_2,
\label{subs2}
\end{eqnarray}
A change of variables like the one above will not always map the independent, integer-valued variables $Q_1$, $J_2$ to new independent, integer valued variables. If $a,b,c,d$ are allowed to be any integer, $G_1$ and $F_2$ may not be independent and therefore the action in terms of these variables will not be equivalent to our original action. However, if $a,b,c,d$ represent an element of the modular group, then the above substitutions represent a valid change of variables. Note that for the duality transform $(a,b,c,d)=(0,-1,1,0)$, and so this transformation gives us $G_1=Q_1$, $F_2=J_2$. 

After performing the above change of variables, we arrive at the following action:
\begin{eqnarray}
&&S[\vec{G}_1,\vec{F}_2]=\frac{1}{2}\sum_k \frac{(2\pi)^2}{|f_k|}\Big[\frac{1}{\tilde{g}}|\vec{F}_2(k)|^2\label{SFG2}\nonumber\\
&&+\left(\tilde{g}+\frac{\tilde{\eta}^2}{\tilde{g}}\right)|\vec{G}_1(k)|^2-\frac{2\tilde{\eta}}{\tilde{g}}\vec{F}_2(-k) \cdot \vec{G}_1(k)\Big],
\label{SFG}
\end{eqnarray}
where 
\begin{eqnarray}
\tilde{g}&=&\frac{g}{(d+\eta c)^2+g^2c^2},\label{gtilde}\\
\tilde{\eta}&=&\frac{(b +\eta a)( d+\eta c)+g^2ca}{(d+\eta c)^2+g^2c^2}.
\label{etatilde}
\end{eqnarray}

In the above we have specialized to the potential given in Eq.~(\ref{V}), on the symmetry line where $v_1=v_2$. One can check that the new parameters $\tilde{\eta}$, $\tilde{g}$ are precisely those given by Eq.~(\ref{zstar}).
We can then dualize the $F_2$ to obtain the following action:
\begin{eqnarray}
S[\vec{G}_1,\vec{G}_2]&=&\frac{1}{2}\sum_k \frac{2\pi\tilde{g}}{|\vec{f_k}|}\left[ |\vec{G}_1(k)|^2+|\vec{G}_2(k)|^2\right]\nonumber\\
&+&\sum_k i2\pi\tilde{\eta} \vec{G}_1(-k) \cdot \vec{a}_{G_2}(k),
\label{SG}
\end{eqnarray}
with $\vec{G}_2=\vec{\nabla}\times\vec{a}_{G_2}$. We can now understand the gapped variables as linear combinations of $J$ variables and $Q$ variables. 
A more general version of the above equation is given in the Appendix. 

In the above we chose a modular operation to map a given $\eta,g$ to the region of the phase diagram where $\tilde{g}^2+\tilde{\eta}^2>1,|\tilde{\eta}|\leq\frac{1}{2}$. We could instead have chosen a different operation which mapped to $\tilde{g}^2+(\tilde{\eta}-n)^2>1$, $|\tilde{\eta}-n|\leq\frac{1}{2}$, since this would still be in the $\infty$ parameter region. The coefficients of this new transformation would change by $a\rightarrow a+nc$, $b\rightarrow b+nd$, with $c$ and $d$ remaining unchanged. We can see from Eq.~(\ref{subs2}) that the new variables $G_1$ (and by symmetry $G_2$) are the same regardless of which part of region $\infty$ the original model is being mapped to. In what follows we will always choose the modular operation which maps to the region $\tilde{g}^2+\tilde{\eta}^2>1,|\tilde{\eta}|\leq\frac{1}{2}$. We have found that all physical results depend only on the coefficients $c$ and $d$, which are the same regardless of which part of the $\infty$ parameter region the $\tilde{g},\tilde{\eta}$ variables are in.

\begin{figure}[t]
\includegraphics[angle=-90,width=\linewidth]{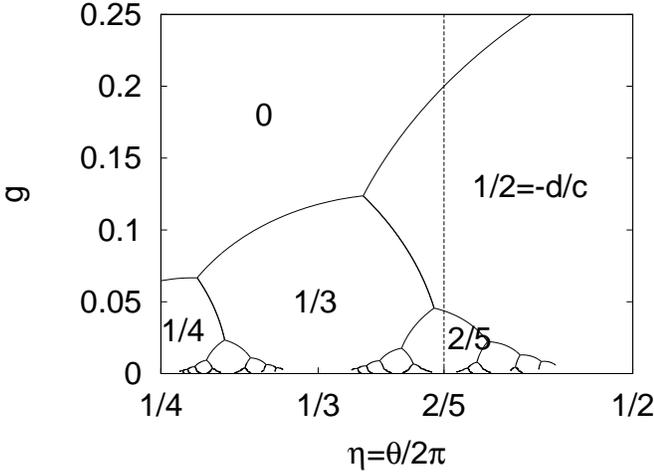}
\caption{A section of Fig.~\ref{SL2Z}, blown up to show the labels on the various phases. Not all transitions are shown as there are infinitely many of them at small $g$. The dashed line is at $\eta=2/5$, where the data in Figs.~\ref{C11}, \ref{C12}, and \ref{sigmaxy} were taken. } 
\label{labels}
\end{figure}

We will label each phase by the label $\frac{-d}{c}$. (Note that for phases with $0<\eta<1$, $c$ and $d$ have opposite signs). For any modular transformation, $c$ and $d$ are mutually prime, so this label will be an irreducible fraction which uniquely identifies $c$ and $d$. This label is practical for a number of reasons. From Eq.~(\ref{subs2}) we see that it gives the nature of the gapped quasiparticles in this phase. It also gives the $\eta$ value at which this phase touches the $g=0$ axis, which is also the $\eta$ value which maps to $g=\infty$. Phase $\infty$ with $c=0$, $d=1$, and phase 0 with $c=1$, $d=0$ both conform to this label. Figure \ref{labels} shows a section of the modular invariant phase diagram with the labels assigned. 

We can understand each phase as a condensate of objects which have $G_1=0$ or $G_2=0$. An example of such an object would have $Q_1=d$, $J_2=c$. In our Monte Carlo simulations we can greatly reduce autocorrelation times by attempting worms of these composite objects.

Let us provide some examples of the application of the above approach. Consider performing an experiment on this system by decreasing $g$ while holding $\eta$ constant at $\eta=\frac{1}{n}$, with $n$ an integer, and $n\neq2$. This is equivalent to a vertical sweep in Fig.~\ref{SL2Z}, or a sweep along the symmetric line in a figure similar to Fig.~\ref{phase}. At large $g$, the $J$ variables are gapped. As $g$ is decreased the $J$ variables condense and the $Q$ variables are gapped. In fact, the precise meaning of condensation of the $J$ variables is that their dual $Q$ variables are gapped.\cite{short_range3} Though the intraspecies potential of the $J$ particles (which is controlled by $g$) is small, the $J$ variables feel a statistical interaction, and this limits how many loops of the $J$ variables can form. Also, $g$ is large enough that large values of $J$ are still costly, and so we expect that when the $J$ variables first condense they have strength one. This means that complex composite objects of the $J$ variables are not condensed here. Consider further decreasing $g$. In Eq.~(\ref{gstar}), we see that for non-zero $\eta$, at small enough $g$ the parameter $g_{\rm dual}$ will also become small. Now we enter a phase where both the $Q$ and $J$ variables want to proliferate in some form. The gapped variables are now the $G$ variables given by the modular operation $ST^{n}S$, which has coefficients $(a,b,c,d)=(-1,0,n,-1)$, leading to 
\begin{equation}
\tilde{g}=\frac{1}{n^2g}, \quad \tilde{\eta}=-\frac{1}{n}.
\end{equation}
As $g\rightarrow 0$, these variables will have $\tilde{g}\rightarrow\infty$, and therefore no loops. The small value of $g$ thus leads to the binding and condensation of more complicated composites of $J$ (in particular $Q_1=1$, $J_2=-n$). These objects will see no statistical interaction, and loops of these variables can form more easily than loops of the $J$ variables in phase 0. Specifically, under the change of variables in Eqs.~(\ref{subs1}) and (\ref{subs2}), the interactions in Eq.~(\ref{SFG}) are such that the $G_1$ variables want to be gapped and the $F_2$ variables condensed (hence the $G_2$ variables are also gapped). This is phase ``$1/n$''. The transition from phase 0 to phase $1/n$ occurs at $g=1/n(\sqrt{n^2-1})$.

Now consider the same experiment as above, this time holding $\eta=\frac{2}{5}$, as shown by the vertical line in Fig.~\ref{labels}. Phase $\infty$ and phase 0 have the same properties as in the earlier case. At $g=\frac{1}{5}$, the system enters phase 1/2. The new gapped variables in this phase are related to the $J$ variables by the operation $ST^{2}S$, and so have $\vec{G}_1=2\vec{Q}_1+\vec{J}_2$. 
At $g=1/(5\sqrt{21})$ the system enters phase 2/5. The new gapped variables correspond to the operation $ST^{-2}ST^{2}S$, which has $(a,b,c,d)=(-2,1,-5,2)$. They remain gapped even as $g\rightarrow 0$. The new condensed variables see no statistical interaction and can condense completely.

In the general case of rational $\eta$ such that $\eta=\frac{r}{s}$, with $r$ and $s$ mutually prime integers, we can find a modular transformation such that $c=s$, $d=-r$. For such a transformation, we can see from Eq.~(\ref{gtilde}) that $\tilde{g}=1/(gc^2)\rightarrow \infty$ as $g\rightarrow 0$. Therefore for general $\eta$ the system will pass through a number of phases with different gapped variables, before finally reaching a phase on the $g=0$ axis where $c$ and $d$ are related to $\eta$, and which can be viewed as a condensate of composite objects like $Q_1=r$, $J_2=s$.

\section{Correlation functions and Conductivities}
\label{sec:correlations}
In our Monte Carlo, we can measure the correlators between the current variables, $C_J^{11}(k)$ and $C_J^{12}(k)$, where the $J$ subscript refers to the fact that these are correlators in the $J$ variables. Here and below we are dropping the direction subscripts on these variables: $C^{11}$ means $C^{11}_{xx}$ and $C^{12}$ means $C^{12}_{xy}$. We would like to determine the values of these correlators in the thermodynamic limit for all the phases in the phase diagram. The conductivities $\sigma^{11}_{xx}(k), \sigma^{12}_{xy}(k)$ are functions of these correlators, so this will also give us the values of these conductivities. We know the values of the correlators in phase $\infty$, because in this phase the $J$ variables are gapped. This means that the only excitations are small loops, which implies that $C_J^{11}(k)\sim k^2$. Since we measured the correlators at $k=(0,0,k_z=\frac{2\pi}{L})\equiv k_{\rm min}$, we find that $C^{11}_J(k_{\rm min})\sim \frac{1}{L^2}$. The smallest excitation that contributes to $C^{12}_J(k)$ consists of a small loop in each of the $\vec{J}_1$ and $\vec{J}_2$ variables. An estimate of such contributions with cross-linking between the loops leads to $C^{12}_J(k_{\rm min}) \sim -\sin(\theta)k_{\rm min}^3\sim1/L^3$. From these correlators we can determine that the conductivities vanish in this phase. 

\begin{figure}[t]
\includegraphics[angle=-90,width=\linewidth]{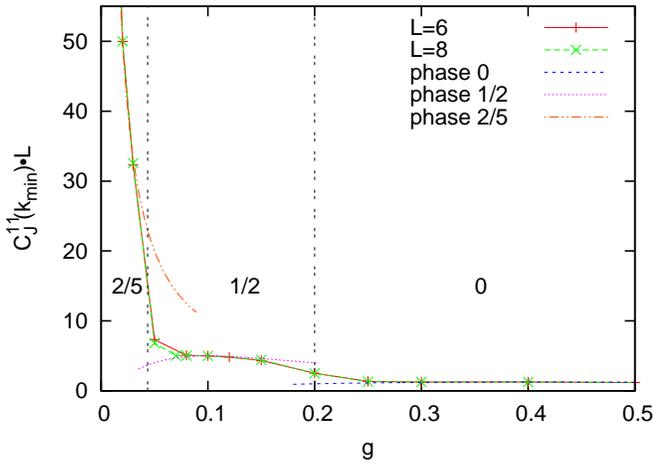}
\caption{$C_J^{11}(k_{\rm min})\cdot L$ as a function of $g$ for $\eta=\frac{2}{5}$ (i.e.~along the dashed line in Fig.~\ref{labels}), for system sizes $L=6$ and $L=8$. The dashed curves correspond to the theoretical predictions in the different phases. The dotted vertical lines denote the phase transitions, which occur at $g=\frac{1}{5}$ and $g=\frac{1}{5\sqrt{21}}$. } 
\label{C11}
\end{figure}

\begin{figure}[t]
\includegraphics[angle=-90,width=\linewidth]{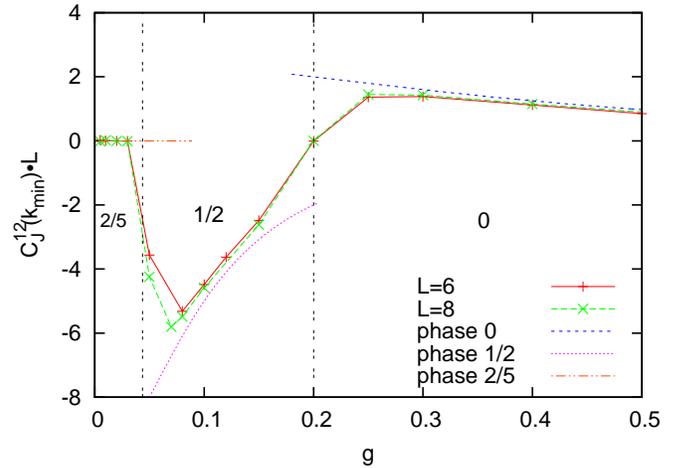}
\caption{$C_J^{12}(k_{\rm min}) \cdot L$ as a function of $g$ for the same system as in Fig.~\ref{C11}. The dashed curves correspond to the theoretical predictions. The dotted vertical lines denote the phase transitions.} 
\label{C12}
\end{figure}

In the previous section, we have interpreted each phase by going to the appropriate $G_1$ and $G_2$ variables, and since these variables are gapped in this phase, we know the behavior of the $G$ correlators for the same reasons given above. Therefore we wish to express $C_J^{11}(k)$ and $C_J^{12}(k)$ in terms of $C_G^{11}(k)$ and $C_G^{12}(k)$, where the latter are correlators of the new variables. To do this, consider the combination 
\begin{equation}
D_J(k)=\pi\left[\frac{C^{12}_J(k)}{\sin{\frac{k_z}{2}}} +\frac{i C^{11}_J(k)}{|\sin{\frac{k_z}{2}}|}\right]. 
\label{Ddef}
\end{equation}
Consider the effect of the duality operator $S$ on this object. We can derive the following relation between the complex correlation $D_J(k)$ in the direct variables, and the complex correlation $D_Q(k)$ in the dual variables:\cite{short_range3}
\begin{equation}
D_Q(k)=z^2D_J(k)+z.
\label{Qrelation}
\end{equation}
Note that this equation is a relation between two \emph{different} correlation functions at the \emph{same} point in the phase diagram. The periodicity operator $T$ does not change the correlation functions. Combining the actions of the two operators leads to the following relation between correlation functions $D_G(k)$ of the $G$ variables in Eq.~(\ref{Ddef}), and the correlation functions $D_J(k)$ of the original $J$ variables:\cite{Fradkin_SL2Z}
\begin{equation}
D_G(k)=(cz+d)^2 D_J(k)+c(cz+d),
\label{Dz}
\end{equation}
where $c$ and $d$ are the parameters of the modular group operation which gives the gapped quasiparticles. We can rewrite Eq.~(\ref{Dz}) to get expressions for the $J$ correlation functions in terms of only the $G$ variables:
\begin{widetext}
\begin{equation}
C_J^{11}(k)=\frac{|\sin{\frac{k_z}{2}}|}{\pi}\frac{c^2g}{(c\eta+d)^2+c^2g^2}+
\frac{[(c\eta+d)^2-c^2g^2]C_G^{11}(k)-2cg(c\eta+d)\sgn{(k_z)}C_G^{12}(k)}{[(c\eta+d)^2+c^2g^2]^2},
\label{C11J}
\end{equation}
\begin{equation}
C_J^{12}(k)=\frac{\sin{\frac{k_z}{2}}}{\pi}\frac{-c(c\eta+d)}{(c\eta+d)^2+c^2g^2}+
\frac{[(c\eta+d)^2-c^2g^2]C_G^{12}(k)+2cg(c\eta+d)\sgn{(k_z)}C_G^{11}(k)}{[(c\eta+d)^2+c^2g^2]^2}.
\label{C12J}
\end{equation}
\end{widetext}
In the phase where $G$ are gapped, $C_G^{11}(k_{\rm min})\sim 1/L^2$ and $C_G^{12}(k_{\rm min})\sim 1/L^3$, and so in the thermodynamic limit the behavior of $C_J^{11}(k_{\rm min})$ and $C_J^{12}(k_{\rm min})$ is given by the first terms in the above expressions. Our numerical results agree with this analysis. Plots of these first terms compared to the numerical data are given in Fig.~\ref{C11} for $C_J^{11}(k_{\rm min})$, and Fig.~\ref{C12} for $C_J^{12}(k_{\rm min})$. To find the curves that correspond to the theoretical predictions, one reads off $c$ and $d$ from the label of a given phase, and substitutes these coefficients into the leading terms of Eqs.~(\ref{C11J}) and (\ref{C12J}). 

\begin{figure}[t]
\includegraphics[angle=-90,width=\linewidth]{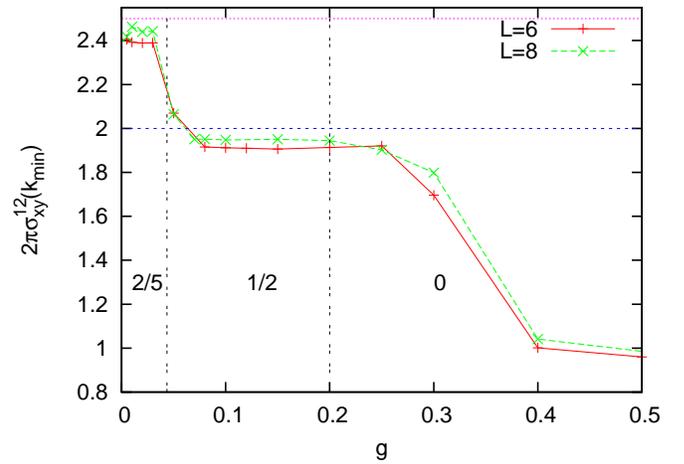}
\caption{$2\pi\sigma^{12}_{xy}(k_{\rm min})$ as a function of $g$ for the same system as in Fig.~\ref{C11}. We observe that the conductivity approaches $\frac{1}{2\pi}2$ in phase 1/2 and $\frac{1}{2\pi}\frac{5}{2}$ in phase 2/5, as expected. The dotted vertical lines denote the phase transitions. In phase 0, $\sigma^{12}_{xy}$ approaches non-universal values.} 
\label{sigmaxy}
\end{figure}

From the correlation functions we can also determine the conductivities. We find that in the thermodynamic limit (and assuming $d\neq0$)
\begin{eqnarray}
\sigma^{11}_{xx}(k_{\rm min})&=&0, \\
\sigma^{12}_{xy}(k_{\rm min})&=&\frac{-c}{2\pi d}.
\label{sigma}
\end{eqnarray}
These conductivities are determined solely by the coefficients $c$ and $d$, and hence by the ``$-d/c$'' label. Figure \ref{sigmaxy} shows $\sigma^{12}_{xy}(k_{\rm min})$ for $\eta=\frac{2}{5}$.  As $g$ is decreased, the system passes through both phase 1/2 and 2/5, and the conductivity takes the expected values of $-\frac{1}{2\pi}\frac{c}{d}=\frac{1}{2\pi}2$ and $\frac{1}{2\pi}\frac{5}{2}$ in these phases. At a phase transition the $G$ variables are not gapped and so the above expressions do not hold. In phase 0, $d=0$, and so to determine the behavior of the conductivities one must take into account the subleading terms in Eqs.~(\ref{C11J}) and (\ref{C12J}). When this is done one finds that $\sigma_{xx}^{11}(k_{\rm min})$ diverges in phase 0 (which is why its crossings can be used to detect phase transitions) and $\sigma_{xy}^{12}(k_{\rm min})$ approaches a non-universal value.

Note that in the above expressions, if we chose a different modular operation which had $\tilde{\eta}\rightarrow\tilde{\eta}+n$, this would not change $D_G(k)$, nor would it change the modular group coefficients $c$ and $d$, and therefore the above equations would remain unchanged. Therefore shifting $\tilde{\eta}$ by an integer, which is equivalent to choosing a different modular operation to describe the gapped particles, does not change any of the physical properties. 

A different situation arises when we discuss shifting $\eta$ by an integer, i.e.~$\eta\rightarrow\eta+n$. This corresponds to describing a different point on the phase diagram, for example a point with $\eta\approx1/3$ after a shift of 1 would have $\eta\approx4/3$. From Eq.~(\ref{action}), it is clear that the correlators $C_J^{11}(k)$, $C_{J}^{12}(k)$ should have the same properties after the shift, but this is not obvious from Eqs.~(\ref{C11J}) and (\ref{C12J}). However, in order to get an action in terms of gapped quasiparticles in a phase at the shifted $\eta$, we must apply a $T^{-n}$ operation to our action before we apply the modular operation for unshifted $\eta$. This changes the modular coefficients $b$ and $d$: $b\rightarrow b-an$, $d\rightarrow d-cn$, and this cancels the shift in $\eta$ in Eqs.~(\ref{C11J}) and (\ref{C12J}) to leave the correlators unchanged. Therefore when $\eta$ is shifted different quasiparticles become gapped. 

Though the current-current correlators do not change when $\eta$ is shifted, the conductivities do change [note the dependence on $d$ in Eq.~(\ref{sigma})]. Though the equivalence of the correlators implies that the system's response to an applied field is unchanged by a shift in $\eta$, this does not mean that the system's response to the internal fields is unchanged. In particular,\cite{short_range3} when defining the conductivities we are treating the statistical interaction as a long-ranged interaction mediated by real-valued internal gauge fields. Shifting $\eta$ changes the strength of this interaction, which in turn changes the action of the internal gauge fields. This is responsible for the change in the conductivity. An interesting case is the effect of such a shift on the conductivity in phase 0. In this phase the $J$ variables are condensed, and since these are the variables which carry the current in this phase $\sigma^{11}_{xx}$ diverges, as we have seen. We can apply the operator $T^1$ to phase 0 to get phase 1. In this phase, the partition function for the $J$ variables is exactly the same, but $\sigma^{11}_{xx}$ does not diverge. To understand this, recall the precise meaning of condensation: a variable is said to be condensed if its dual variables under the formal duality transformation are gapped. The variables dual to the $J$ variables are the $Q$ variables that are gapped in phase 0. However, the $Q$ variables are not gapped in phase 1 and hence the $J$ variables are not condensed in the above sense. Instead, some other variables, which can be determined from the substitutions in Eqs.~(\ref{subs1}) and (\ref{subs2}) appropriate for phase 1, are condensed. Another way of interpreting condensation is that in calculations like the current-current correlations we can replace integer-valued condensed variables by real-valued variables, and perform Gaussian integrals over these variables. By the above arguments we can do this in phase 0 but not phase 1, and noting how we defined the conductivity for the $J$ currents, it ``knows'' whether or not this real-valued replacement is possible. This explains the difference in conductivities between phase 0 and phase 1.

The phase diagram in Fig.~\ref{SL2Z} has a number of special ``fixed'' points which are unchanged by an operation of an element of the modular group. There are two types of such points: ``triple points'' where three phase transitions meet, and points halfway along a phase transition line, such as the point at $\eta=0,g=1$.\cite{Fradkin_SL2Z,Lutken1993} The invariance under a modular operation means that in Eqs.~(\ref{C11J}) and (\ref{C12J}) we have $C_J^{11}(k)=C_G^{11}(k)$ and $C_J^{12}(k)=C_G^{12}(k)$. We can then solve the two equations to determine the correlation functions, and therefore also the conductivities. We obtain the following results, applicable at all fixed points with $g>0$:
\begin{eqnarray}
C^{11}_J(k)&=&\frac{|\sin(\frac{k_z}{2})|}{2\pi g},\\
C^{12}_J(k)&=&0. 
\label{fixedC}
\end{eqnarray}
We can then determine the conductivities:
\begin{eqnarray}
\sigma^{11}_{xx}(k)&=&\frac{g}{2\pi(g^2+\eta^2)}, \\
\sigma^{12}_{xy}(k)&=&\frac{\eta}{2\pi(g^2+\eta^2)}.
\end{eqnarray}
We have verified these equations numerically for the following points $(\eta,g)$: $(1/2,\sqrt{3}/2), (1/2,1/2), (1/2,\sqrt{3}/6)$. When $\eta=0$ the two species of particles are decoupled, and we studied this system in Ref.~\onlinecite{Ranged_Loops}. We found the above equations to hold for the fixed point $(\eta=0,g=1)$.

\section{Nature of transitions}
\label{sec:histograms}
\begin{figure}[t]
\includegraphics[angle=-90,width=\linewidth]{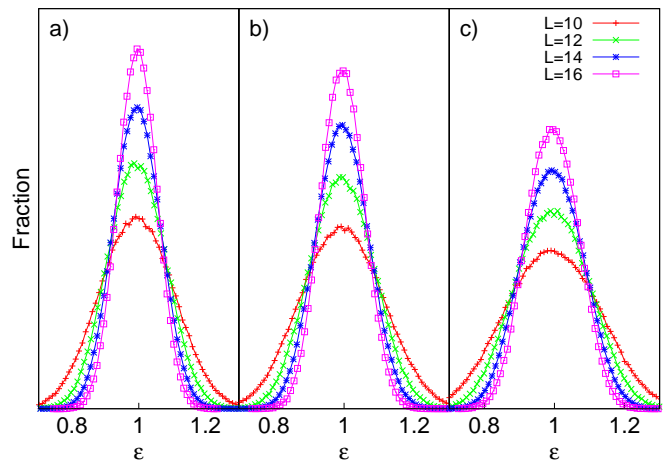}
\caption{Histograms of $\epsilon$ at various points on the phase transition between phase $\infty$ and phase 0, using sizes $L=10,12,14,16$. A normally-distributed histogram implies that the transition is second-order, while a transition with multiple peaks implies first-order. The first panel corresponds to $\eta=\frac{1}{5}$, the next two panels correspond to $\eta=\frac{1}{4}$,$\frac{1}{3}$. We see no evidence of first-order behavior at these sizes.} 
\label{histcirc}
\end{figure}

Our numerical approach allows us to study the properties of the various phase transitions in Fig.~\ref{SL2Z}. We have attempted to determine the order of the transition between phase $\infty$ and phase 0. To do this we study histograms of the total energy $\epsilon$ at the phase transition, utilizing the fact that we know the exact location of the transition. In a second-order transition, we would expect such histograms to be normally distributed, while for a first-order transition we may see multiple peaks in the distribution. Figure \ref{histcirc} shows histograms taken on the $\infty$-0 transition, at $\eta=\frac{1}{5},\frac{1}{4},\frac{1}{3}$. Histograms for system sizes $L=10,12,14$ and $16$ are shown. We see normally distributed histograms, suggesting a second-order transition. We can show that in our sign-free reformulation using the $Q_1$, $J_2$ variables, $\langle\epsilon\rangle=1-\frac{1}{L^3}$ for the model with $J_1\leftrightarrow J_2$ interchange symmetry, at all values of $g$ and $\eta$. Our Monte Carlo measurements of $\langle\epsilon\rangle$ confirm this.  

\begin{figure}[t]
\includegraphics[angle=-90,width=\linewidth]{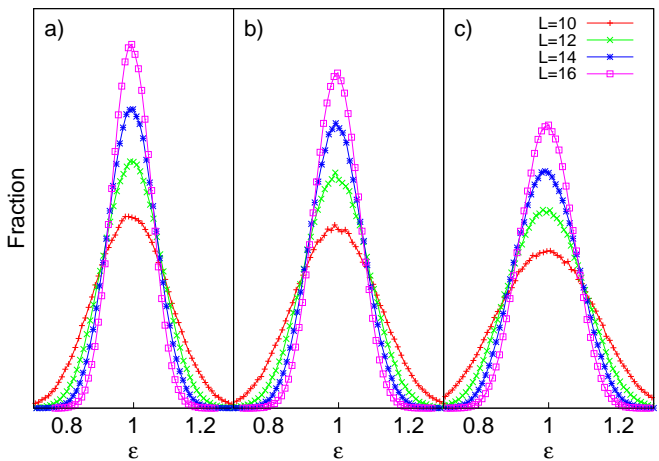}
\caption{Histograms of $\epsilon$ at various points on line of phase transitions at $\eta=\frac{1}{2}$, using sizes $L=10,12,14,16$. Each point is related to a point on the $\infty$-0 phase transition by the operation $T^1ST^1S$. The first panel maps to $\eta=\frac{1}{5}$, the next two panels map to $\eta=\frac{1}{4}$,$\frac{1}{3}$. We see no evidence of first-order behavior. The histograms are also identical to those in Fig.~\ref{histcirc}, which provides a check on our Monte Carlo.} 
\label{histtop}
\end{figure}

\begin{figure}[t]
\includegraphics[angle=-90,width=\linewidth]{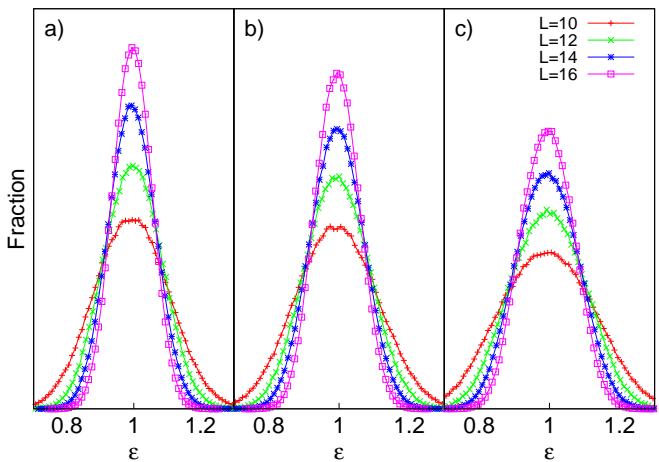}
\caption{Same as Fig.~\ref{histtop}, but the operation to map these points to $\infty$-0 phase transition is $ST^{-1}S$, as explained in the text.} 
\label{histbottom}
\end{figure}

The modular invariance of the system implies that all phase transitions that are related by a modular operation will have the same properties. In fact, one can show that in our variables $Q_1,J_2$, any two points related by the modular group produce simulations with the same energies, so the histograms should be identical. In these variables the updates used in the Monte Carlo are different, but if they are done properly the results should be the same. We will check this by studying the properties of the line of phase transitions at $\eta=\frac{1}{2}$. There are two modular group operations which map the $\infty$-0 phase transition to this one. The first is $T^1ST^1S$, which has $(a,b,c,d)=(0,-1,1,-1)$. This maps the three above points at $\eta=\frac{1}{5},\frac{1}{4},\frac{1}{3}$ to three points with $\eta=\frac{1}{2}$ and $g=\sqrt{6}/4,\sqrt{15}/6,1/\sqrt{2}$. Histograms at these points are shown in Fig.~\ref{histtop}. Once again, we see no evidence of first-order behavior at these system sizes. The second modular group operation which maps the $\infty$-0 transition to the line at $\eta=\frac{1}{2}$ is $ST^{-1}S$, which has $(a,b,c,d)=(-1,0,-1,-1)$. This maps the three points on the semi-circle to $\eta=\frac{1}{2}$ and $g=1/\sqrt{6},\sqrt{15}/10,1/(2\sqrt{2})$. Histograms for these points are shown in Fig.~\ref{histbottom}, and they also show no sign of first-order behavior. The histograms for the related points in Figs.~\ref{histcirc}, \ref{histtop} and \ref{histbottom} are identical, as predicted by the above argument. 


\begin{figure}[t]
\includegraphics[angle=-90,width=\linewidth]{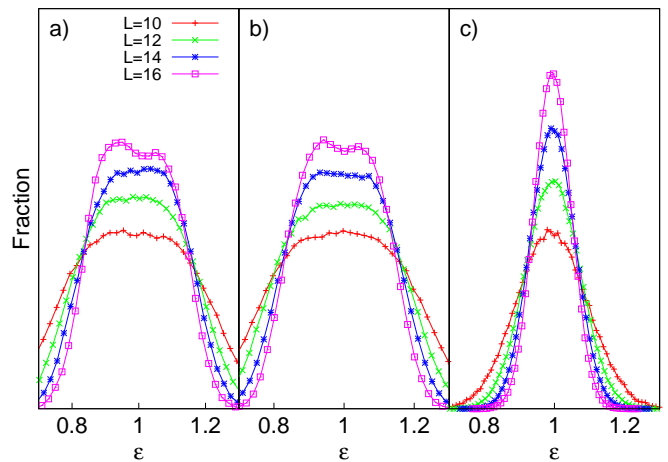}
\caption{Histograms of $\epsilon$ at some special ``fixed'' points in Fig.~\ref{SL2Z} using sizes $L=10,12,14,16$. The first two panels show histograms at ``triple points'' where three phase transitions meet. Panel (a) shows the triple point at the upper end of the line of phase transitions at $\eta=\frac{1}{2}$, which has $g=\sqrt{3}/2$. Panel (b) shows the point at the lower end with $g=\sqrt{3}/6$. Both histograms have a double-peaked structure which indicates that the transitions at these points are first-order. Panel (c) shows histograms at the fixed point $\eta=\frac{1}{2}, g=\frac{1}{2}$. These histograms show no sign of first-order behavior, which we expect since this point should have the same properties as the point $\eta=0,g=1$ which is known to be continuous.\cite{Ranged_Loops}}
\label{histspec}
\end{figure}

We have also studied the system at the ``triple points'' on the modular invariant phase diagram, where three phase transitions meet. We expect all such points to have the same properties, and we have studied the points at the ends of the $\eta=\frac{1}{2}$ line of phase transitions, which occur at $g=\sqrt{3}/2$ and $g=\sqrt{3}/6$. Histograms for these points are shown in Fig.~\ref{histspec}(a) and (b). We see that the histograms have two clear peaks, indicating that these are first-order transitions. 

In Ref.~\onlinecite{Ranged_Loops} we studied the phase transition at $(\eta=0,g=1)$ and found it to be continuous. This point maps to the point $\eta=\frac{1}{2},g=\frac{1}{2}$, and we have studied the phase transition at this point to confirm the second-order behavior. Histograms at this point are shown in Fig.~\ref{histspec}(c), and we see no sign of first-order behavior.

\section{Critical exponents of putative second-order transitions}
\label{sec:exponents}
\begin{figure}[t]
\includegraphics[angle=-90,width=\linewidth]{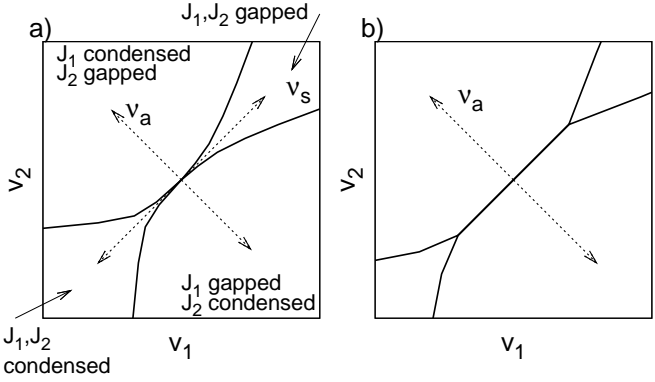}
\caption{(a) Schematic blow-up of the phase diagram in the $v_1$, $v_2$ (Fig.~\ref{phase}) variables, near a tetracritical point. At such a point we expect two scaling directions with distinct critical exponents $\nu_s$ and $\nu_a$. Due to the symmetry on interchange of $J_1$ and $J_2$, we expect the scaling directions to be in the symmetric and antisymmetric directions, shown by the dotted lines. The shape of the phase transition lines implies that $\nu_s > \nu_a$. (b) Phase diagram for the special case of $\theta=\pi$ $(\eta=1/2)$. We can see that along the symmetric direction the system goes along the phase boundary, and we can only drive a phase transition across the antisymmetric direction. } 
\label{tetra}
\end{figure}

Apart from the triple points, the transitions we have studied are second-order, and we can now determine their critical exponents. In Fig.~\ref{SL2Z} the $\infty$-0 transition seems to take place at an ordinary critical point, however in Fig.~\ref{phase} we see that the transition actually takes place at a tetracritical point when we allow $v_1$ and $v_2$ to be not equal. Figure \ref{tetra}(a) shows a closer look at the upper tetracritical point in Fig.~\ref{phase}. At such a tetracritical point there are two scaling directions, each with a different correlation length exponent.\cite{FisherRGreview} Our system is symmetric under the interchange of $J_1$ and $J_2$, which implies that one scaling direction is in the symmetric ($\delta v_1=\delta v_2$) direction, with critical exponent $\nu_s$, and the other is in the antisymmetric direction ($\delta v_1=-\delta v_2$), with critical exponent $\nu_a$. We can determine which of these exponents is larger based on how the phase transition lines meet.
A simple renormalization group argument shows that the phase boundaries in the local coordinates $\lambda_s,\lambda_a$ obey
\begin{equation}
 \lambda_a \sim \lambda_s^{\nu_s/\nu_a}.
\end{equation}
Our data shows that the phase transition lines have the same shape as those in Fig.~\ref{tetra}, and this combined with the above equation implies $\nu_s>\nu_a$. 

\begin{figure}[t]
\includegraphics[angle=-90,width=\linewidth]{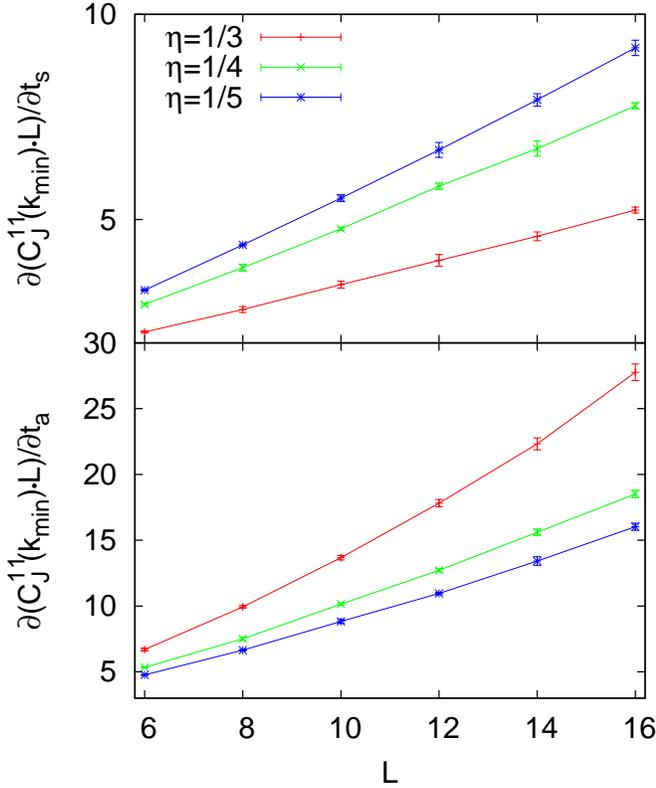}
\caption{Plots of the derivatives $\partial (C^{11}_J(k_{\rm min})\cdot L)/\partial t_{s}$ and $\partial (C^{11}_J(k_{\rm min})\cdot L)/\partial t_{a}$ at various points on the $\infty$-0 phase transition. Error bars were obtained by comparing the results of independent simulations. We expect such plots to scale as $L^{1/\nu_{s,a}}$. The values shown in Fig.~\ref{bignu} were extracted by fitting these curves to the function $a+bL^{1/\nu}$. }
\label{drhoeven}
\end{figure}

\begin{figure}[t]
\includegraphics[angle=-90,width=\linewidth]{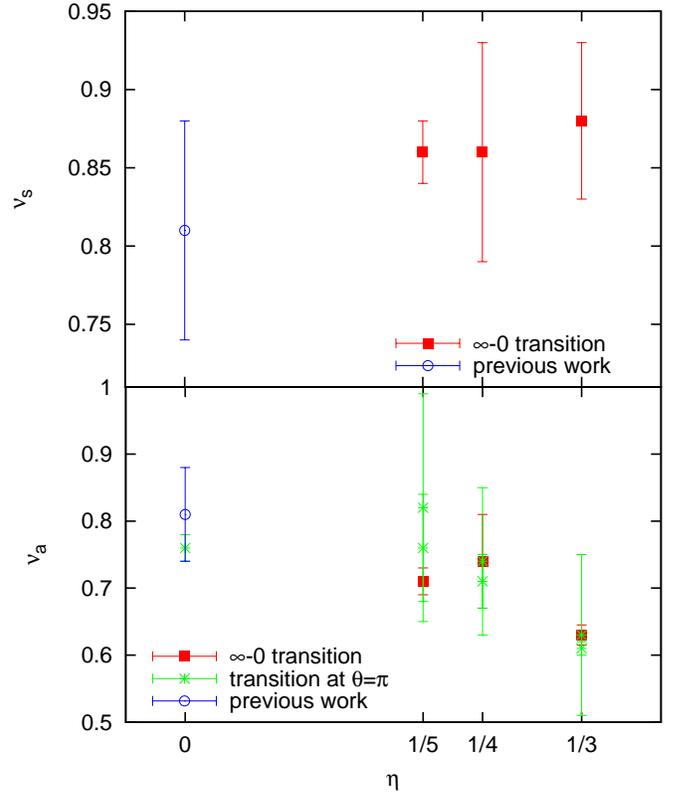}
\caption{Critical exponents along the (a) symmetric and (b) antisymmetric scaling direction, extracted from the data in Fig.~\ref{drhoeven}. Error bars come from the fits, and are further increased to account for finite-size effects as discussed in the text. The blue circles represent the value of $\nu$ obtained in Ref.~\onlinecite{Ranged_Loops}. $\nu_s$ cannot be measured when $\theta=\pi$, since the system does not cross a phase transition in this direction [see Fig.~\ref{tetra} (b)]. On the $\theta=\pi$ line there are two points which map to each point on the $\infty$-0 semicircle.} 
\label{bignu}
\end{figure}

We will extract the critical exponents by taking appropriate derivatives of $C^{11}(k_{\rm min})\cdot L$. In this system with long-ranged interactions, $C^{11}(k_{\rm min})\cdot L$ approaches a constant (possibly zero) value in each phase. At a critical point, it jumps from one value to another, which leads to a peak in its derivative. As mentioned in Sec.~\ref{sec:MC}, we want to take derivatives with respect to a short-ranged part of the potentials, given by the parameters $t_1$ and $t_2$ in Eq.~(\ref{Vt}). To extract $\nu_s$ we take derivatives with respect to the symmetric combination $t_s$, while for $\nu_a$ we use the antisymmetric combination $t_a$. $C^{11}(k_{\rm min})\cdot L$ has the scaling form:
\begin{eqnarray*}
&&C^{11}(k_{\rm min})\cdot L=f_a(L t_s^{\nu_s}) \quad \textrm{symmetric direction},\\
&&C^{11}(k_{\rm min})\cdot L=f_s(L t_a^{\nu_a}) \quad \textrm{antisymmetric direction},
\end{eqnarray*}
where $f_a$ and $f_s$ are scaling functions. This leads to
\begin{eqnarray}
&&\left.\frac{\partial C^{11}_J(k_{\rm min})\cdot L}{\partial t_s}\right|_{t_s=t_a=0}\sim L^{1/\nu_{s}} \label{scaling1},\\
&&\left.\frac{\partial C^{11}_J(k_{\rm min})\cdot L}{\partial t_a}\right|_{t_s=t_a=0}\sim L^{1/\nu_{a}} \label{scaling2},
\label{scaling}
\end{eqnarray}
so $\nu_{s,a}$ can be extracted by fitting curves of $\partial (C^{11}_J(k_{\rm min})\cdot L)/\partial t_{s,a}$ vs $L$. Such curves, at the $\infty$-0 transition, are shown in Fig.~\ref{drhoeven}(a) for the symmetric derivative and Fig.~\ref{drhoeven}(b) for the antisymmetric derivative. The extracted values of $\nu_s$ and $\nu_a$ are shown in Fig.~\ref{bignu}. We see that $\nu_s > \nu_a$, as expected from the shapes of the phase transition lines near the tetracritical point. We also see that neither exponent is close to $1/3$, supporting the conclusion that we have second-order transitions at these points. The exponent $\nu_a$ is decreasing as we move along the phase transition away from $\eta=0$. On the other hand, the error bars on $\nu_s$ are too large to tell whether it is varying. Error bars for $\nu$ can be estimated from the fits to the $\partial (C^{11}_J(k_{\rm min})\cdot L)/\partial t$ curves. However, we may have significant finite size effects in our results, even though we are simulating exactly at the transition, because we do not know the subleading corrections to Eqs.~(\ref{scaling1}) and (\ref{scaling2}). To account for this in the error bars we performed fits both including and not including the data at $L=6$, and if the errors from the fitting procedure were not large enough to encompass both values we increased the error bars. The values of $\nu$ were taken from the fits which included the $L=6$ data. We have also plotted the critical exponents measured in Ref.~\onlinecite{Ranged_Loops} at the point $g=1,\eta=0$. At this point the two species of particles are decoupled, so $\nu_s=\nu_a$. 

We could not determine $\nu_s$ on the $\eta=\frac{1}{2}$ line because changing $t_s$ does not move the system through a phase transition, instead it moves along the line of phase transitions, as seen in Fig.~\ref{tetra}(b). On the other hand, we can argue that the transition driven by $t_a$ is equivalent to that on the $\infty$-0 line at the related point, and the $\nu_a$ values on this transition are shown in Fig.~\ref{bignu}(b).

The transition at $\eta=\frac{1}{2}$ is a transition between phase I, where the $J_1$ variables are condensed and the $J_2$ are gapped, and phase II where $J_1$ is gapped and $J_2$ condensed. The $\pi$-statistical interactions prevent both types of loops from condensing simultaneously. The two species could behave as immiscible fluids and phase separate, or they could coexist in a critical soup. \cite{Senthil2006_theta,shortlight,deccp_science} Our result that the transition is second-order implies that the two species can indeed form such a critical state. 

An open question is how three transitions meet at a triple point in the modular phase diagram in Fig.~\ref{SL2Z}. All three transitions could be second-order all the way to the triple point, or they could have bicritical points where they become first-order. Our results show that on at least part of the phase boundaries the transition is second-order. A more detailed study could determine whether the phase transitions do become first-order at some point, and where this point is.

\section{Discussion}
We studied a model of two species of particles with mutual statistics and long-ranged interactions such that the model has the same form after the application of operations from the modular group. Using this modular invariance, we were able to analytically conjecture the phase diagram and determine the values of the current-current correlations and conductivities in each phase and at all points that are invariant under the modular group. We can also describe each phase in terms of particles gapped in that phase. Using a reformulation of the model that does not have the sign problem, we performed Monte Carlo studies and firmly established the conjectured phase diagram. Furthermore, we numerically determined the order of the transitions in the phase diagram of this modular invariant system. We found the triple points to be first-order but the other phase transitions we studied were second order. The second-order transitions are evidence for a novel critical loop soup state in the case $\theta=\pi$. 

Exact results derived from the modular invariance greatly improved our numerical study. We were able to derive a number of useful checks on the Monte Carlo, as well as run simulations exactly at the location of the critical points, greatly simplifying our measurements of critical exponents. 

Since we have only studied this model at relatively small system sizes, it is possible that the transitions which we observe to be continuous are actually first-order. Studying larger system sizes could confirm the second-order behavior. It would also be interesting to study field theories for such systems with marginally long-ranged interactions and mutual statistics.\cite{Ranged_Loops,Radzihovsky1995,Kuklov2005,short_range3}
Studying this model at larger system sizes would also allow one to obtain better estimates of $\nu$, since any subleading terms would have a reduced effect. In addition, at larger sizes one could determine the behavior near the triple points, in particular where the transition changes from first-order to second-order. However, the long-ranged nature of the interactions in the model means that in our reformulation, the amount of computer time needed scales as $L^6$, making such larger studies difficult, but possible with future resources.

At $\theta=\pi$, our (2+1)-dimensional model is relevant to the study of unusual phase transitions in (2+1) dimensions.\cite{Senthil2006_theta,shortlight,deccp_science} It also applies to the study of (2+1)-dimensional symmetry protected topological (SPT) phases and phase transitions,\cite{LevinStern2009,Lu2012,Senthil2012} as well as the surface states of (3+1)-dimensional SPT phases.\cite{Vishwanath2012} It may also be possible to use similar lattice models to study such SPT phases in the bulk. \cite{Chen2011,Chen2011b,Neupert2011,ChoDyon,Swingle2012,Xu2012,Keyserlingk2012} More generally, our loop model provides a precise lattice realization of a topological field theory. It would be interesting to study such lattice models for other topological field theories.\cite{Barkeshli2010,Kou2009,Wen2000,Hansson2004,Cho2011,Xu2011,Burnell2011}

\acknowledgments
We are thankful to A.~Vishwanath, M.~P.~A.~Fisher, I.~Gruzberg, A.~Kitaev, N.~Read, T.~Senthil and S.~Simon for stimulating discussions. This research is supported by the National Science Foundation through grants DMR-0907145 and DMR-1206096; by the Caltech Institute of Quantum Information and Matter, an NSF Physics Frontiers Center with support of the Gordon and Betty Moore Foundation; and by the XSEDE computational initiative grant TG-DMR110052.

\appendix
\section{Modular transformations for general potentials}
The method in Sec.~\ref{sec:reformulations} allows us to apply duality and periodicity operations to an action to obtain a new action in terms of new variables. Such a procedure is possible for any choice of potentials, though potentials other than the ones used in this work will not be modular invariant. We can still interpret the new variables as being gapped in a certain phase, but without modular invariance we cannot use this to determine the exact locations of the phase transitions, or to predict where the new variables will be gapped. In Ref.~\onlinecite{short_range3} we used numerical methods to determine which variables were gapped in each phase, and we were then able to use this procedure to find the action for these gapped variables. We were also able to determine the correlation functions and conductivities in the phases where we knew the gapped variables, using the same methods as in Sec.~\ref{sec:correlations}.

We now provide the equations that generalize the methods we have used in this paper to any potential. These equations also represent the procedure used in Ref.~\onlinecite{short_range3}, generalized to any operation of the modular group. The action in terms of new variables is obtained by starting with Eq.~(\ref{Snum}), making the substitutions $Q_1,J_2\rightarrow G_1,F_2$ given in Eqs.~(\ref{subs1}) and (\ref{subs2}), and then dualizing the $F_2$ variables to obtain the $G_2$ variables. We find
\begin{eqnarray}
S&=&\frac{1}{2}\sum_k \left[ v_{G_1}(k)|\vec{G}_1(k)|^2+v_{G_2}(k)|\vec{G}_2(k)|^2\right]\nonumber\\
&+&\sum_k i\theta_G \vec{G}_1(-k) \cdot \vec{a}_{G_2}(k),
\end{eqnarray}
where
\begin{eqnarray*}
v_{G_{1/2}}(k)&=&\frac{(2\pi)^2v_{2/1}(k)}{(2\pi d+\theta c)^2+v_1(k)v_2(k)|\vec{f}_k|^2c^2},
\end{eqnarray*}
and
\begin{eqnarray*}
\theta_{G}&=&2\pi \cdot \frac{(2\pi b +\theta a)(2\pi d+\theta c)+v_1(k)v_2(k)|\vec{f}_k|^2ca}{(2\pi d+\theta c)^2+v_1(k)v_2(k)|\vec{f}_k|^2c^2}.\nonumber
\end{eqnarray*}
We can also express the current-current correlators in terms of correlators in the new variables, using the same methods as in Ref.~\onlinecite{short_range3}. For simplicity, we specialize to the symmetric line where $v_1=v_2\equiv v$:
\begin{widetext}
\begin{eqnarray*}
C_J^{11}(k)=\frac{v(k)|f_k|^2c^2}{(\theta c+2\pi d)^2+|f_k|^2v(k)^2c^2}+
\frac{[(\theta c+2\pi d)^2-|f_k|^2v(k)^2c^2]C_G^{11}(k)-4\sin{\frac{k_z}{2}}v(k)c(\theta c+2\pi d)C_G^{12}(k)}{[(\theta c+2\pi d)^2+|f_k|^2v(k)^2c^2]^2}\cdot (2\pi)^2,\\
C_J^{12}(k)=\frac{-2\sin{\frac{k_z}{2}}c(\theta c+2\pi d)}{(\theta c+2\pi d)^2+|f_k|^2v(k)^2c^2}+
\frac{[(\theta c+2\pi d)^2-|f_k|^2v(k)^2c^2]C_G^{12}(k)+4\sin{\frac{k_z}{2}}v(k)c(\theta c+2\pi d)C_G^{11}(k)}{[(\theta c+2\pi d)^2+|f_k|^2v(k)^2c^2]^2}\cdot (2\pi)^2.
\end{eqnarray*}
\end{widetext}
When $v=\frac{2\pi g}{|f_k|}$, these equations reduce to Eqs.~(\ref{C11J}) and (\ref{C12J}).

\bibliography{bib4twoloops}
\end{document}